# Reversible arithmetic logic unit

Rigui zhou, Yang shi*, Manqun Zhang

Quantum computer requires quantum arithmetic. The sophisticated design of a reversible arithmetic logic unit (reversible ALU) for quantum arithmetic has been investigated in this letter. We provide explicit construction of reversible ALU effecting basic arithmetic operations. By provided the corresponding control unit, the proposed reversible ALU can combine the classical arithmetic and logic operation in a reversible integrated system. This letter provides actual evidence to prove the possibility of the realization of reversible Programmable Logic Device (RPLD) using reversible ALU.

*Introduction:* Although more than half a century has already passed, the well-known von-Neumann model[1] still maintains strong vitality, the implementation technologies of it, such as vacuum tubes, transistors, etc. has already changed rapidly. Reversible logic, which means the input vector states can be always reconstructed from the output vector states uniquely, is one of the emerging important implementation technologies, when the end of Moore's law approaching, the researches on reversible logic provide low power CMOS with necessary theory foundation and design method. A quantum computer is built from a quantum circuit containing wires and elementary quantum gates to carry around and manipulate the quantum information. Reversible logic is a core part of the quantum circuit model[4]. Consequently, if we want to construct quantum arithmetic, it must build its reversible logical components, such as reversible full adder, reversible multiplier and so on. Among these reversible circuits, the controlled addition to a reversible circuit is an essential and necessary transition to form a more complex controlled system in a quantum arithmetic network. This letter will make use of the addition circuit introduced by Kai-Wen Cheng et al [2] to construct reversible arithmetic logic unit (ALU) for quantum arithmetic.

Actually, the design in this letter is not the first reversible version ALU. Michael Kirkedal Thomsen et al [4] also gave an ALU design by generalizing the V-shape design which can achieve five basic arithmetic-logical operations on two n-bit operands, but the structure is so simple that some functions can not even get the right results, for example, we can not calculate the *ADD operation* only using the bitwise exclusive-or of the two n-bit arguments $|A>$ and $|B>$ without considering the carry bit or only by setting the carry bit to the TRUE. The goal of this letter is to build a multi-functional circuit reasonably that conditionally performs one of several possible arithmetic-logical operations on two operands $|A>$ and $|B>$ depending on control input data instructions. The following section is based on the assumption that the readers are familiar with the basics of reversible logic[6].

***Reversible ALU:*** This section presents the sophisticated design of a reversible n-bit ALU based on a controlled quantum full adder proposed by Kai-Wen Cheng et al [2] with a sequential arrangement of operations. The proposed reversible ALU intends to implement the following basic set of operations, we assume the two n-bit arguments as $|A>=|a_1 a_2 a_3 .... a_n>$ and $|B>=|b_1 b_2 b_3 .... b_n>$ : 1. ADD operation: $|A+B>$; 2. SUB operation: $|A-B>$; 3. Negative SUB operation: $|B-A>$; 4. Bitwise exclusive-or operation: $|A \oplus B>$; 5. NOT operation; 6. No-operation; 7. OR operation: $|AorB>$; 8. AND operation: $|A \bullet B>$. Later we will see that the final reversible ALU allows more than these eight operations. Although the operations 7 and 8 are clearly irreversible, in the proposed reversible ALU, the irreversibility of these two operations can be eliminated by setting control input data instructions.

**ADD operation:** Let $|a_i>$ be the ith qubit of augend number $|A>$, $|b_i>$ be the ith qubit of addend number $|B>$, $|C_i>$ be the ith qubit of carry, and $|S_i>$ be the ith qubit of sum, some equations can be obtained as followed [3]:

$$C_i = a_i b_i + (a_i \oplus b_i) c_{i-1} = a_i b_i \oplus (a_i \oplus b_i) c_{i-1} \tag{1}$$

$$S_i = a_i \oplus b_i \oplus c_{i-1} \tag{2}$$

Detailed descriptions of the quantum adders can be found in [2], it is summarized here as Fig.1 for the sake of convenience of further study and for added emphasis.

**SUB operation:** According to [5], if we let $|B_i>$ be the ith qubit of borrow, and $|D_i>$ be the ith qubit of defference $A-B$, then the relationship among $|a_i>, |b_i>, |B_{i-1}>$ and $|B_i>$, $|D_i>$ will be:

$$B_i = b_i B_{i-1} \oplus \overline{a_i}(b_i \oplus B_{i-1}) \tag{3}$$

$$D_i = a_i \oplus b_i \oplus B_{i-1} \tag{4}$$

By changing (3), (4) into the following form:

$$B_i = b_i B_{i-1} \oplus \overline{a_i}(b_i \oplus B_{i-1}) = b_i \overline{a_i} \oplus (b_i \oplus \overline{a_i}) B_{i-1} \tag{5}$$

$$D_i = a_i \oplus b_i \oplus B_{i-1} = \overline{\overline{a_i} \oplus b_i \oplus B_{i-1}} \tag{6}$$

According to the equations (5), (6), it can be seen that the *SUB operation* can be replaced by the *ADD operation* by using $|\overline{A}>$、$|B>$ as the augend number and the addend number, and using the $|B_{i-1}>$ as the carry bit, meanwhile, giving a bitwise negation to the sum bit. The bitwise negations can be performed on the adder's input $|A>$ and output $|S>$ by giving two arrays of Feyman gate in the quantum full adder and setting the control lines $C_{SUB}$ and $C_{snot}$ of these Feyman gates in effect simultaneously(Please see the following Figure 2). The NO-operation, which means the identity function and used when instructions in a computing system require that the operands remain unchanged by the ALU[4], can also be composed by setting the control line $C_{SUB}$ FALSE

**Negative SUB operation:** when calculating the $|B-A>$ in a conventional ALU, we have the equality $|B-A>=|B+\overline{A}+1>$, but it is not a suitable approach for a reversible ALU. This letter will make use of the modular subtraction provided by Michael Kirkedal Thomsen et al [4], the definition of modular subtraction was $|B-A>=|\overline{\overline{B}+A}>$, it is obvious that this definition requires two bitwise negations and one addition. We can also design this operation based on the same method as the SUB operation. So this doesn't need much elaboration.

**Bitwise exclusive-or operation:** Actually, if we set the second exclusive-or with $c_{i-1}$ of the sum bit $S_i = a_i \oplus b_i \oplus c_{i-1}$ ineffective, the Bitwise exclusive-or operation will be calculated. But in order to control this operation in the reversible ALU, the control line $C_{carryxor}$ should be added, when all the control bit, including $C_{carryxor}$, $C_{SUB}$ and $C_{snot}$ are FALSE, the extended reversible ALU will perform Bitwise exclusive-or operation of $|A \oplus B>$. At the same time, in order to realize the *NOT operation*, only control input data instruction for the gate that eliminates the effect of the carry bit are needed in the reversible ALU which is shown in Figure 3 with control line $C_{aANDb}$.

**OR operation/AND operation:** The OR operation/AND operation are not elementary operations in reversible logic, we should employ the special reversible logic gates, such as Toffoli gate or Peres gate to realize these particular operations, however, by changing the control lines of the reversible ALU, the proposed ALU can manage changes of the carry bit $C_i$, and so get the results of the OR operation/AND operation. The final reversible ALU which has five control lines is shown in Figure 3. The table 1 shows the selected operations resulting from the combinations of the control lines for the reversible ALU. However, what needs to emphasize is the number of control lines. When looking from the outside, having five control lines to select among these nine operations may seem redundant, but we want to explain here that the purpose of designing the proposed ALU is to achieve the basic arithmetic and logic operation in a reversible way. Actually, when we extend the combinations of the control input data instructions in the proposed design, the extra operations like the bitwise inclusive-OR, increment operations and so on can be easily got.

*Conclusions:* Arithmetic Logic Unit (ALU) for the Programmable Logic Device (RPLD) has been presented in this letter in a reversible way. The reversible version ALU aims at eliminating the energy dissipation effecting by the irreversible action of classical computer. All the suggested operations provided by the proposed ALU are self-inverse or have an easy inverse operation. Also from another perspective, the realization of an efficient reversible ALU shows that a reversible programmable computing device is possible. But it is still a question that the general quantum operations, such as Hadamard, phase shift and rotation by $\pi/4$, etc, can not be integrated into the proposed reversible ALU until now.

*Acknowledgements:* This work is supported by the National Natural Science Foundation of China under Grant No.60873069, the Natural Science Foundation of Jiangxi Province under Grant NO.2009GZS0013 and the Open Project Program of Key Laboratory of Intelligent Computing & Information Processing of Ministry of Education, Xiangtan University, China (No.2009ICIP04).

**Authors' affiliations:**
College of Information Engineering, East China Jiao Tong University, Nanchang, Jiangxi, 330013, China
E-mail: shy2400@gmail.com

Fig.1 The n-bit quantum adder in [2]

Fig.2 SUB operation by setting control lines to the quantum full adder

Fig.3 The sophisticated design of a reversible arithmetic logic unit (reversible ALU)

Table 1 selected operations resulting from the combinations of control lines for the reversible ALU

Figure 1:

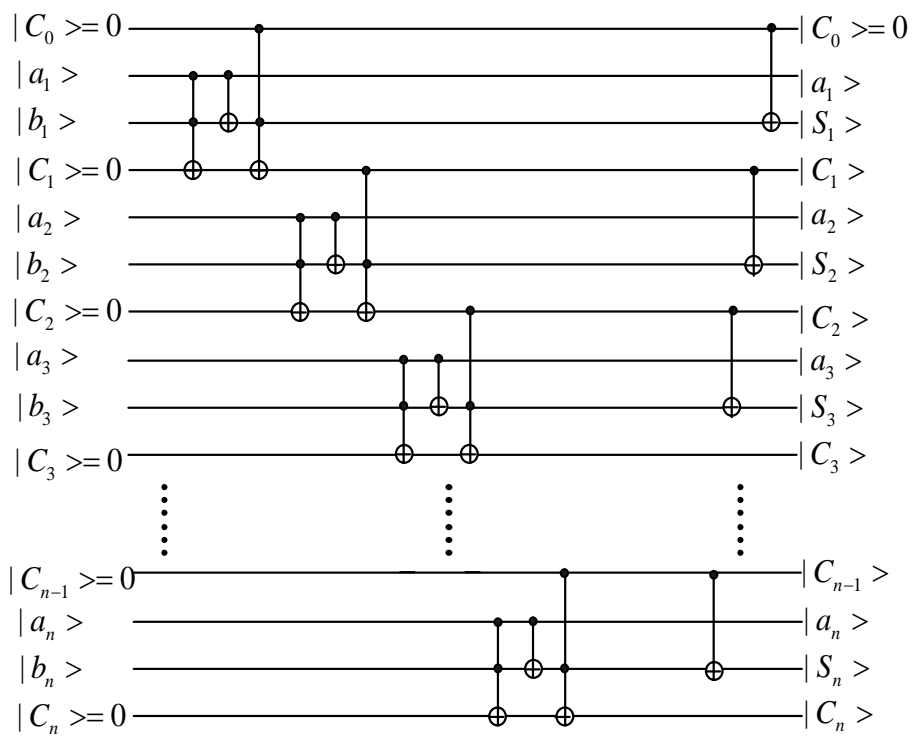

Figure 2:

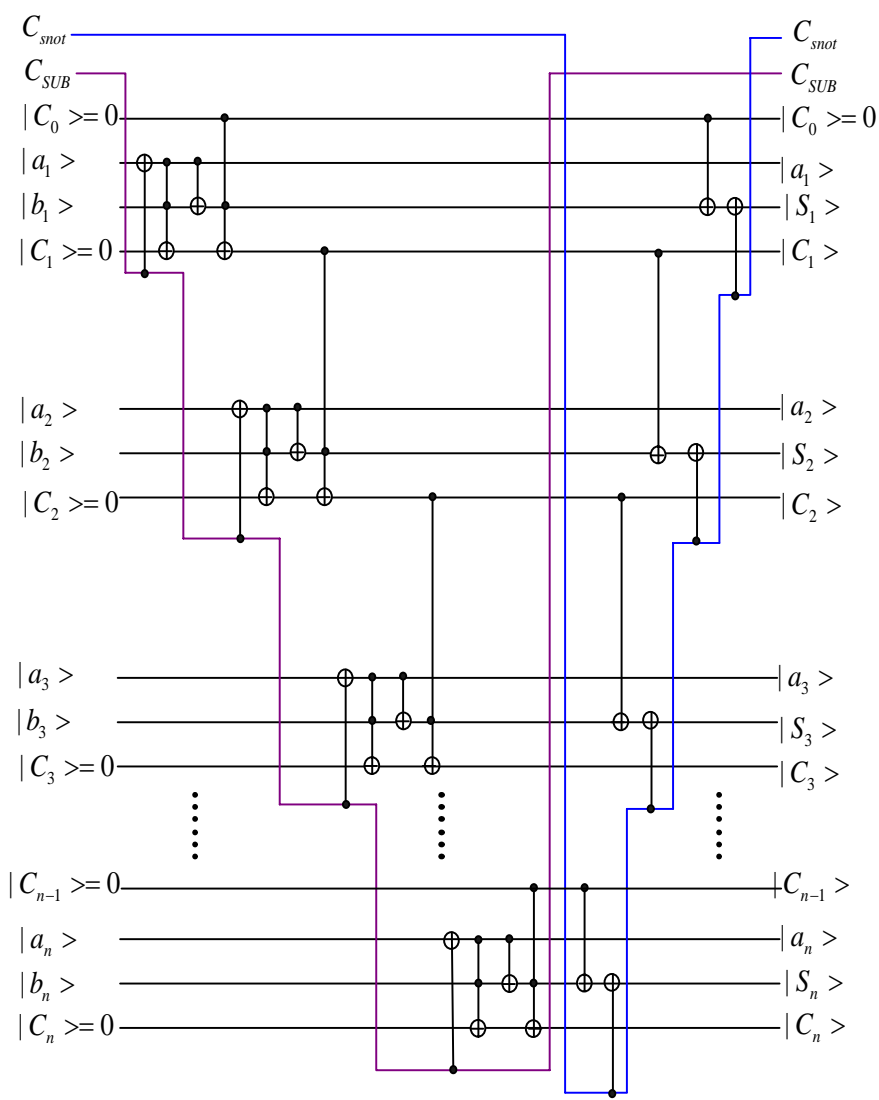

Figure 3:

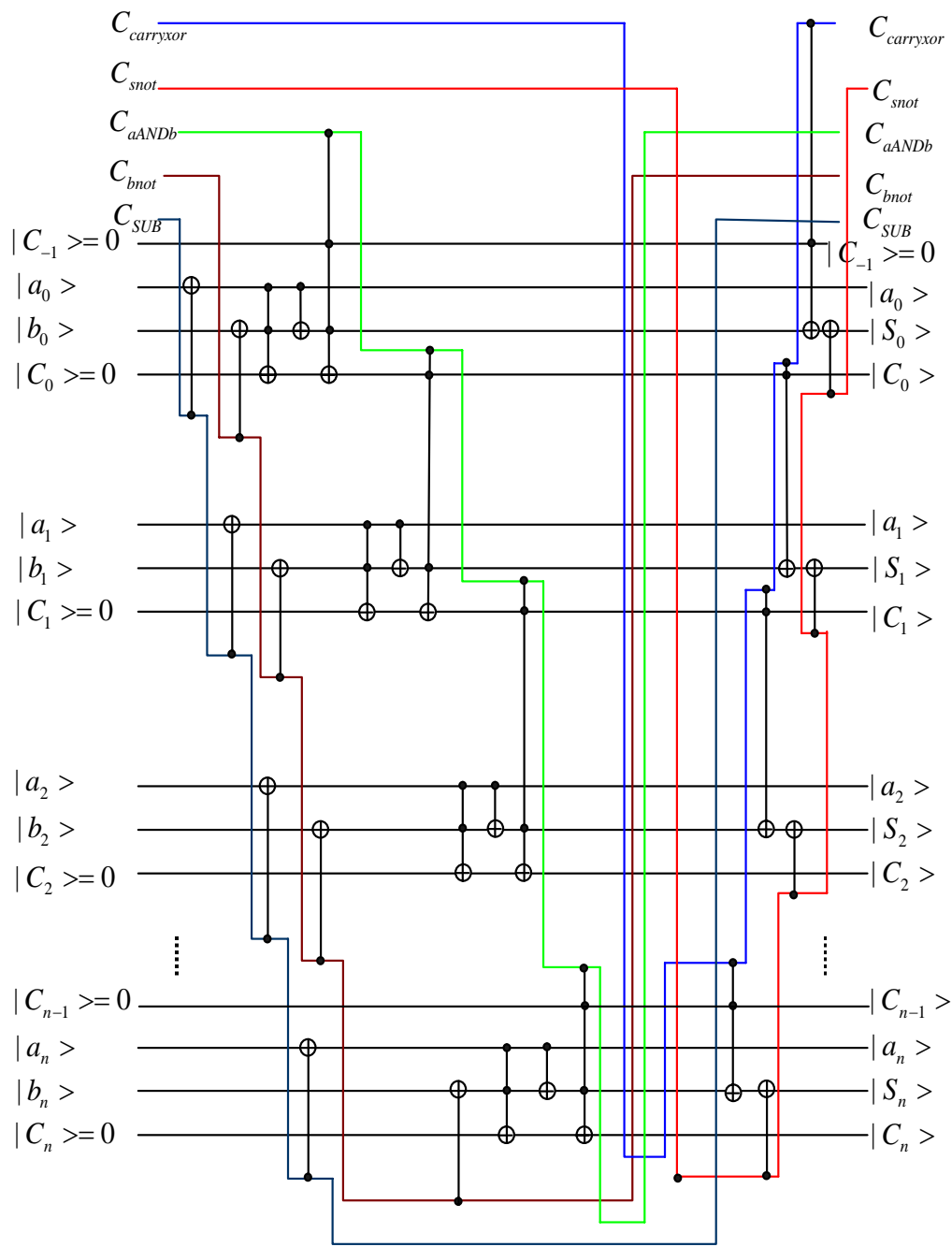

Table 1:

| | | $C_{carryxor}$ | $C_{snot}$ | $C_{aANDb}$ | $C_{bnot}$ | $C_{SUB}$ | operation |
|---|---|---|---|---|---|---|---|
| | (1) | 1 | 0 | 1 | 0 | 0 | $A$加$B$ |
| | (2) | 1 | 1 | 1 | 0 | 0 | $\overline{\overline{A}加B}$ |
| | (3) | 0 | 0 | 0 | 0 | 0 | $A \oplus B$ and $A \cdot B$ |
| | (4) | 0 | 1 | 0 | 0 | 0 | $\overline{A \oplus B}$ |
| | (5) | 0 | 0 | 0 | 0 | 1 | $\overline{A}$ |
| | (6) | 1 | 1 | 1 | 0 | 1 | $A$减$B$ ($A$加$\overline{B}$加 1) |
| | (7) | 1 | 1 | 1 | 1 | 0 | $B$减$A$ ($\overline{\overline{B}加A}$) |
| | (8) | X | X | X | X | X | $NOP$ |
| | (9) | 0 | 1 | 0 | 1 | 1 | $\overline{A} \cdot \overline{B}$ ($\overline{A+B}$) |